\documentclass[%
 aip,
 amsmath,amssymb,
 reprint,%
]{revtex4-1}

\usepackage{graphicx}
\usepackage{dcolumn}
\usepackage{bm}

\usepackage[utf8]{inputenc}
\usepackage[T1]{fontenc}
\usepackage{mathptmx}

\begin{document}

\preprint{AIP/123-QED}

\title[Time-irreversibility tests for random-length time series]{Time-irreversibility tests for random-length  time series: the matching-time approach applied to DNA}

\author{R. Salgado-Garc{\'\i}a}
\email{raulsg@uaem.mx}
\affiliation{Centro de Investigaci\'on en Ciencias-IICBA, Physics Department, Universidad Aut\'onoma del Estado de Morelos. Avenida Universidad 1001, colonia Chamilpa, CP 62209, Cuernavaca Morelos, Mexico. }

\date{\today}

\begin{abstract}
In this work we implement the so-called matching time estimators for estimating the entropy rate as well as the entropy production rate for symbolic sequences. These estimators are based on recurrence properties of the system, which have been shown to be appropriate to test irreversibility specially when the sequences have large correlations or memory. 
Based on limit theorems for matching-times we  derive a maximum likelihood  estimator for entropy rate assuming that we have a set of moderately short symbolic time-series of finite random duration. We show that the proposed estimator has several properties that makes it adequate to estimate entropy rate and entropy production rate (or to test irreversibility)  when the  sample sequences have different lengths such as the coding sequences of DNA. We test our approach in some controlled examples of Markov chains. We also implement our estimators in genomic sequences to show that the degree of irreversibility coding sequences of human DNA is significantly  larger than the corresponding non-coding sequences. 
\end{abstract}

\maketitle

\begin{quotation}

Time irreversibility of a time series gives statistical information about the nature of the underlying process such as the presence of nonlinear correlations~\cite{daw2000symbolic}. Although this concept has been mainly  used in physics to determine how much a given process is far from thermodynamic equilibrium~\cite{latora1999kolmogorov,PorpoAl,RolP,gaspard2004time}, it has been adapted and implemented in other disciplines to study the time irreversibility of a wide range of different phenomena, such as for instance, the heart beat time series~\cite{costa2005broken}, DNA sequences~\cite{salgado2021estimating,Provata2014}, financial time series of assets~\cite{flanagan2016irreversibility} or even pieces of classical music~\cite{Gustavo}. The problem of determining  whether or not is a series time irreversible is not a trivial task due to the finite duration of the measurements (or the time series). Furthermore, in many cases such time series might be even of random duration. The DNA coding sequences, music scores and even heart beat time seres are some examples of this situation. In this case, estimating the degree of irreversibility requiere a different approach from the standard tools already developed. Here we propose a method that profits of the random duration of the time-series when we have access to a sample set of different-length time series. \end{quotation}

\section{\label{sec:intro}Introduction}


Real life measurements of certain variables results in time-series of finite duration. When  there are no restrictions in the number of sample date being acquired the time series collected can have the same length and these can be used to perform statistical analysis without worrying about the statistical errors in each time-series. This is because the error is proportional to the inverse of the square root of sample size. However, time series collected from a certain process are not always of the same length, and even might be of random duration. This is the case, for example, of DNA coding sequences, since the genes codifying for proteins have different lengths~\cite{li1991fundamentals}. Another examples of this situation is the case of symbolic time series coming from  musical pieces~\cite{Gustavo}, the heart beat time series~\cite{costa2005broken} or any other biological signal whose measurements might be abruptly interrupted by uncontrolled external or internal influences. It is clear that the statistical errors in these cases are different due to randomness in the duration of the series and even this phenomenon might cause a bias if the estimator is dependent on the length of the time-series as in the case of the recurrence-time statistics~\cite{Kon,ChR,ChU,cesar2015fluctuations,salgado2021estimating}. 

In this work we are mainly interested in determine the degree of irreversibility of a process by analyzing a sample set of random-length time series, i.e. finite-time realizations of random duration of the corresponding  process. This is made through the use of the recurrence-time statistics~\cite{Kon,ChR,cesar2015fluctuations,salgado2021estimating} and in particular of the matching-time estimator for entropy rate~\cite{Kon}. This estimator has been less studied in practical applications and in this work we show that it is a suitable tool for analyzing this class of inhomogeneous time series. To estimate the entropy rate and \emph{reversed entropy rate} (i.e., the relative entropy of the process with respect to the invariant measure of the time-reversed process~\cite{ChR}) we assume that the matching-time obtained from every time series is normally distributed, i.e., we assume the validity of the central limit theorem for finite, but sufficiently large, time series. This hypothesis allows us to implement the maximum likelihood method to obtain an estimator for a sample set of random-length time series. Thereafter  it is possible to determine the degree of irreversibility of the process by comparing the entropy rate and the reversed entropy rate estimated by this approach. 

This work is organized as follows. In Sec.~\ref{sec:Matching-time} we resume the statistical properties of the matching-time estimator for entropy rate, such as the central limit theorem which was proved in Ref.~\onlinecite{Kon}. In this section we also introduce the maximum likelihood estimator for the entropy rate when considering as sample set of random length time series. 
In Sec.~\ref{sec:numerical} we test the proposed estimator in two different scenarios: in the case in which the symbolic sequences generated from a reversible Markov chain and from an irreversible Markov chain. In both cases we obtain symbolic sequences of finite random duration using a given distribution of lengths. In Sec.~\ref{sec:testing-DNA} we implement the proposed estimator to determine the degree of irreversibility of coding DNA sequences of human genome, in which every gene is seen as a symbolic time series of random duration. Finally in Sec.~\ref{sec:conclusions} we summarize the main results of our study as well as the main conclusions of this work.

\section{\label{sec:Matching-time}Matching-time estimators for entropy rate and entropy production rate}


\subsection{\label{ssec:matching}Matching times}


Let $\mathbf{X}  := \{ X_{n} \, : \, n\in \mathbb{m}_0\}$ be a discrete-valued stationary ergodic process generated by the law $\mathbb{P}$, whose realizations are infinite sequences of symbols taken from a finite set $A$, i.e., the set of all posible realizations is a subset of $A^{\mathbb{N}}$. Here we denote by $\mathbf{x} = (x_{0} x_{1} x_{2} x_{3}\dots)$ an infinite realization of the process $\mathbf{X}$. Let  $\ell$ be a positive integer, we denote by $x_{0}^{\ell-1}$ the string of the first $\ell$ symbols of the realization $\mathbf{x}$. A finite string $\mathbf{a} := a_1 a_2a_3\ldots a_\ell$ comprised of $\ell$ symbols will be called either $\ell$-word or $\ell$-block. We say that the $\ell$-word $\mathbf{a}$ ``occurs'' at the $k$th site of the trajectory $\mathbf{x}$, if $x_k^{k+\ell-1} = \mathbf{a}$. 

The \emph{entropy rate} $h$ of the process $\mathbf{X}$ (also called Kolmogorov-Sinai entropy or KS entropy) is defined as rate at which the entropy of $n$-blocks grows with $n\in \mathbb{N}_0$~\cite{Kon,ChR}, 
\begin{equation}
h := \lim_{n\to \infty} \frac{1}{n} \sum_{x_0^n\in {A}^n} \mathbb{P} (X_0^{n-1}=x_0^{n-1})\log(\mathbb{P} (X_0^{n-1}=x_0^{n-1})).
\end{equation}
Analogously, the \emph{reversed entropy rate}  $h_\mathrm{R}$ of the process $\mathcal{X}$ is defined as~\cite{Kon,ChR},
\begin{equation}
h_\mathrm{R} := \lim_{n\to \infty} \frac{1}{n} \sum_{x_0^{n-1}\in {A}^n} \mathbb{P} (X_0^{n-1}=x_0^{n-1})\log(\mathbb{P} (X_0^{n-1}=x_{n-1}^0)).
\end{equation}

The \emph{entropy production rate} $e_\mathrm{p}$ of the process $\mathcal{X}$ quantifies the degree of irreversibility of the process $\mathcal{X}$ and is defined as~\cite{Kon,ChR},
\begin{equation}
e_{\mathrm{p}} := \lim_{n\to \infty} \frac{1}{n} \sum_{x_0^{n-1}\in {A}^n} \mathbb{P} (X_0^{n-1}=x_0^{n-1})\log\left( \frac{\mathbb{P} (X_0^{n-1}=x_0^{n-1}) }{\mathbb{P} (X_0^{n-1}=x_{n-1}^0)}\right).
\end{equation}

According to Ref.~\onlinecite{gaspard2004time} the entropy production rate can be obtained as the difference between the reversed entropy rate and the entropy rate, $e_{\mathrm{p}} = h_{\mathrm{R}}-h$,  for Markov processes and for more general systems~\cite{Mae}. This fact allows to state that the difference between the reversed entropy rate and entropy rate, $h_{\mathrm{R}} - h$,  can be used as an irreversibility index in systems in which the entropy production rate cannot be obtained directly from the observed time-series. Now we proceed to state the corresponding estimators for $h$ and $h_{\mathrm{R}}$ based on the so-called \emph{matching times}.

Given a finite realization $\mathbf{x} := x_0 x_1,x_2,\dots x_t$ up to time $t$ of the process $\mathcal{X}$, the \emph{matching time} $L_t^+$ is defined as the shortest $\ell$ such that the $\ell$-word ${x}_0^{\ell-1}$ does not reappears in $\mathbf{x}$. Specifically we have that
\begin{equation}
\label{eq:L+:def}
L_t^+ (\mathbf{x}):= \min\{ \ell \, :\,  {x}_0^{\ell-1}  \not= {x}_{j}^{j+\ell-1},\, \forall j=1,2,\dots, t-\ell+1 \}. 
\end{equation}
Analogously we define the \emph{reversed matching time}  $L_t^-$  as the shortest $\ell$ such that the \emph{time-reversed} $\ell$-word $\mathbf{x}_{\ell-1}^0$ does not appears in $\mathbf{x}$, i.e., 
\begin{equation}
\label{eq:L-:def}
L_t^-(\mathbf{x}) := \min\{ \ell \, :\,  {x}_{\ell-1}^0  \not= {x}_{j}^{j+\ell-1}, j=1,2,\dots, t-\ell+1 \}. 
\end{equation}

According to Kontoyiannis~\cite{Kon}, $L_t^+$ satisfy a law of large numbers, in the sense that $L_t^+/\log(t)$ converges $\mathbb{P}$ almost surely  to $1/h$ for $t\to \infty$, where $h$ is the entropy rate of the process $X$. Analogously the reversed matching times converges  $\mathbb{P}$ almost surely to the reversed entropy rate~\cite{ChR}, i.e., $L_t^-/\log(t) \to 1/h_\mathrm{R}$ for $t\to \infty$. In a similar way, these estimators for the entropy rate and reversed entropy rate also satisfy a central limit theorem in the following form, 
\begin{eqnarray}
\label{eq:cltL+}
\sqrt{ \frac{ h^3 \log(t)}{\sigma^2} }  \left(   \frac{L_t^+}{ \log(t) }  - \frac{1}{h}  \right) &\to& \mathcal{N}(0,1), 
\\
\label{eq:cltL-}
\sqrt{ \frac{ h_R^3 \log(t)}{\sigma_R^2} } \left(   \frac{L_t^-}{ \log(t)} - \frac{1}{h_R}  \right) &\to& \mathcal{N}(0,1), \end{eqnarray}
in distribution as $t\to \infty$. Here $\sigma$ and $\sigma_{\mathrm{R}}$  are a constants depending on the process (see Ref.~\onlinecite{Kon} for details) and in this case can be interpreted as parameters to be estimated related to the statistical error due to the finiteness of the size of the sample sequence. 

Next, we introduce the random variables $X^+$ and $X^-$ as follows
\begin{eqnarray}
X^+_t := \frac{L_t^+}{\log(t)},
\\
X^-_t := \frac{L_t^-}{\log(t)},
\end{eqnarray}
whose realizations can be interpreted as observation of an approximation to the inverse entropy rate and reversed entropy rate, respectively,  for a time series of finite length $t$. 

Notice that both $X^+_t $ and $X^-_t$ satisfies the central limit theorem, as we stated in Eqs.~(\ref{eq:cltL+}) and~(\ref{eq:cltL-}), which allows us to assume that the distribution of  $X^+_t $ and $X^-_t$ is normal for finite, but sufficiently large, $t$. This assumption implies that the probability distribution function of $X_t^+$, which we denote by $f_+ (s)$,  and the probability distribution function of $X_t^-$, denoted by $f_- (s) $, can be approximated by the normal distribution as follows, 
\begin{eqnarray}
\label{eq:dist+}
f_+ (x) &=&  \frac{1}{\sqrt{2\pi \varrho_+^2 }} \exp\left(-\frac{(x-1/h)^2}{2\varrho_+^2} \right),
\\
f_- (x) &=&  \frac{1}{\sqrt{2\pi \varrho_-^2 }} \exp\left(-\frac{(x-1/h_{\mathrm{R}})^2}{2\varrho_-^2} \right),
\label{eq:dist-}
\end{eqnarray}
respectively, for $t\gg 1$. Here $\varrho^2_+$ and $\varrho^2_-$  are the variance of $X_t^+$ and $X_t^-$ respectively,  which depend on the length of sample sequence $t$. Explicitly  $\varrho^2_+$ and $\varrho^2_-$ are defined as
\begin{eqnarray}
\varrho^2_+ &:=& \frac{ \sigma^2 }{h^{3} \log(t)},
\\
\varrho^2_- &:=&   \frac{ \sigma_R^2 }{h_R^{3} \log(t)}.
\end{eqnarray}

Under this assumption, it is clear that the mean and variance of $X_t^+$ are given by
\begin{equation}
\mathbb{E}[X_t^+] = \frac{1}{h},
\qquad
\mbox{Var}(X_t^+) = \varrho^2_+.
\end{equation}
Analogously we also have that
\begin{equation}
\mathbb{E}[X_t^-] = \frac{1}{h_{\mathrm{R}}},
\qquad
\mbox{Var}(X_t^-) = \varrho_{-}^2.
\end{equation}

\subsection{\label{ssec:estimation}{Estimation procedure}}


Now let us state the problem we are facing. We will assume that we have a set $\mathcal{W}:= \{ \mathbf{x}_i, :\, |\mathbf{x}_i| = t_i,   1\leq i \leq m\}$ of $m$ finite sample sequences (finite-time observations of the process $\mathcal{X}$). The sample sequences are assumed to have different lengths $t_i$ and we wish to estimate the entropy rate and the reversed entropy rate.  It is reasonable to assume that the sequence lengths $t_i$ are independent realizations of certain random variable whose distribution is denoted by $ g(t)$. The collection of all the sequence lengths $t_i$  will be denoted by $\mathcal{T}$, i.e., 
\begin{equation}
\mathcal{T} := \{ t_i \, : \, 1\leq i \leq m\}.
\end{equation}

It is clear that each sequence in  $\mathcal{W}$ gives us a sample of the matching time and a sample of the reversed  matching time by applying eq.~(\ref{eq:L+:def}) and eq.~(\ref{eq:L-:def}) respectively. We denote by
\begin{equation}
\mathcal{L}^+ := \{ \ell_i^+ = L_{t_i}^+(\mathbf{x}_i) \, :\, \mathbf{x}_i \in \mathcal{W},\, 1\leq i \leq m \},
\end{equation}
the set of all resulting matching times by applying Eq.~(\ref{eq:L+:def}) to each word in  $\mathcal{W}$. Analogously we denote by 
\begin{equation}
\mathcal{L}^- := \{ \ell_i^- = L_{t_i}^-(\mathbf{x}_i) \, :\, \mathbf{x}_i \in \mathcal{W},\, 1\leq i \leq m \},
\end{equation}
the set of all reversed matching times obtained by applying Eq.~(\ref{eq:L+:def}) to each word in  $\mathcal{W}$. 

The sample sets of matching times, $\mathcal{L}^+$ and  $\mathcal{L}^-$, give in turn \emph{inhomogeneous} sample sets of entropy rate and reversed entropy rate as follows,
\begin{eqnarray}
\mathcal{X}^+ &:=& \left\{ x_i^+ := \frac{\ell_i}{\log(t_i)} \, :\, \ell_i \in \mathcal{L}^+, 1\leq i \leq m  \right\}
\\
\mathcal{X}^- &:=& \left\{ x_i^- := \frac{\ell_i}{\log(t_i)} \, :\, \ell_i \in \mathcal{L}^-, 1\leq i \leq m  \right\}
\end{eqnarray}

We say that these sample sets are \emph{inhomogeneous} in the sense that the collected samples comes from \emph{different} distributions, since the distriibution itself depends on $t$, the length of the time series. In other words, we can think of  $\mathcal{X}^+ $  and  $\mathcal{X}^- $  as a sets of realizations of random variables that are independent but not identically distributed, contrary to what is commonly assumed in statistics. Despite these sample sets are made up of realizations of different distributions, such distributions depend parametrically on the entropy rate and reversed entropy rate respectively, as it can be appreciated in Eqs.~(\ref{eq:dist+}) and~(\ref{eq:dist-}) . This fact allows us to  implement some estimators for $h$ and $h_{\mathrm{R}}$. In Appendix~\ref{ape:ape1} we show that the maximum likelihood estimators for these quantities are given by,
\begin{eqnarray}
\hat{h} &=& \frac{\frac{1}{m} \sum_{j=1}^m \log(t_i) }{\frac{1}{m} \sum_{j=1}^m \ell_i^+ }
\label{eq:hat-h}
\\
\hat{h}_{\mathrm{R}} &=& \frac{\frac{1}{m} \sum_{j=1}^m \log(t_i) }{\frac{1}{m} \sum_{j=1}^m \ell_i^-}
\label{eq:hat-hR}
\end{eqnarray} 
where $\ell_i^+$ and  $\ell_i^-$ are sample matching times from $\mathcal{L}^+$ and $\mathcal{L}^-$  respectively. We should stress the fact that these estimations do not depend on the specific distribution of the sequences length $t_i$, as we proved in Appendix~\ref{ape:ape1}. This allows us to apply our formulas in several possible scenarios in which the sequence lengths are randomly distributed.

The parameters $\sigma^2$ and $\sigma_{\mathrm{R}}^2$ can also be estimated and these quantities allow to obtain the estimated error in the estimation due to the finiteness of the sample sequences. In Appendix~\ref{ape:ape1} we also show that,
\begin{eqnarray}
\hat{\sigma}^2 &=& \hat{h}^2 \left( \hat{h} \hat{c}^+  - \hat{a}^+\right),
\\
\hat{\sigma}^2_{\mathrm{R}} &=& \hat{h}_R^2 \left( \hat{h}_R \hat{c}^-  - \hat{a}^-\right),
\end{eqnarray}
where the sample functions $\hat{a}^{\pm}$ and $\hat{c}^{\pm}$ are defined as
\begin{eqnarray}
\hat{a}^{\pm} := \frac{1}{m} \sum_{i=1}^m \ell_{i}^{\pm}  ,
\\
\hat{c}^{\pm} :=   \frac{1}{m} \sum_{i=1}^m\frac{(\ell^{\pm}_i)^2 }{\log(t_i)}.
\end{eqnarray}

Once we have an expression for  $\sigma^2$ and $\sigma_{\mathrm{R}}^2$, the errors in the estimation due to the finiteness of the sample sequence can be obtained from the central limit theorem through equations~(\ref{eq:cltL+}) and~(\ref{eq:cltL-}). In Appendix~\ref{ape:ape1} we show that the average errors due to the finiteness of the sample sequences are given by,
\begin{eqnarray}
\label{eq:def-error-estimation}
\hat{\varepsilon} = \frac{\hat \sigma^2}{{ \hat h } } \frac{1}{m}\sum_{i=1}^m \frac{1}{{\log(t_i)}},
\\
\label{eq:def-error-R-estimation}
\hat{\varepsilon}_{\mathrm{R}} =\frac{\hat \sigma_R^2}{{ \hat h_R } } \frac{1}{m}\sum_{i=1}^m \frac{1}{{\log(t_i)}}.
\end{eqnarray}
These expressions for $\hat{\varepsilon} $ and $\label{eq:def-error-R-estimation}$ allow us evaluate how large is the error due to the fact that the limit $t\to \infty$ has not been reached and due to the fact the sample sequences are inhomogeneous in their length.

\section{\label{sec:numerical}Numerical experiments}


In this section we will perform numerical simulations in order to test the proposed procedure to obtain estimation for the entropy rate and the reversed entropy rate and  consequently, an estimation for the entropy production rate. We consider the case of a three-states Markov chain which, depending on a parameter, can be reversible or irreversible, thus allowing to test the estimator in both situations.

\subsection{\label{ssec:markov}{Markov chain model}}


To test the estimator for entropy rate we use a three-states Markov chain that is a minimal model of an irreversible stochastic model~\cite{Jiang}. Actually, depending on a parameter, the chain can be reversible or irreversible, thus allowing to test the estimator in both situations, showing that our method allows to estimate  the entropy rate. Consequently, we can use the estimator as an index of irreversibility to apply in real situations.

Let $\{X_t \in \mathbf{S}: t\in \mathbb{N}\}$ be a discrete-time stochastic process with state space $\mathbf{S} := \{1,2,3\} $. We define the process as a three state Markov chain by introducing the stochastic matrix $M : \mathbf{S}\times \mathbf{S} \to [0,1]\subset \mathbb{R}$, defined as
\begin{equation}\label{eq:stochastic}
M = \left(
  \begin{array}{ccc}
   0 & p & 1-p \\
   1-p & 0 & p \\
   p & 1-p & 0
  \end{array} \right),
\end{equation}
where $p$ is a parameter such that $p\in[0,1]$. It is easy to see that this matrix is doubly stochastic and therefore the invariant distribution $\mathbf{\pi} = \mathbf{\pi} M $ is given by $\pi = (\frac{1}{3},\frac{1}{3},\frac{1}{3})$. Moreover, it is easy to compute the entropy rate and the time-reversed entropy rate, which are given by~\cite{Jiang},
\begin{eqnarray}
h(q)&=& - q\log(q)-(1-q)\log(1-q),
\label{eq:h-ex}
\\
h_R(q)&=&-(1-q)\log(q)-q\log(1-q).
\label{eq:hr-ex}
\end{eqnarray}
Additionally, the corresponding entropy production rate is given by
\begin{equation}
e_p(q) = (2q-1)\log\left( \frac{q}{1-q}\right).
\label{eq:ep-ex}
\end{equation}
It is clear from the above formulas that the chain is reversible only for the case $p = 1/2$. Otherwise, the process is irreversible with a degree of irreversibility increasing as $p$ deviates from $p = 1/2$.

\subsection{\label{ssec:fixed-length}{Entropy estimation tests for fixed length time series}}

%
\begin{figure}[t]
\begin{center}
\scalebox{0.35}{\includegraphics{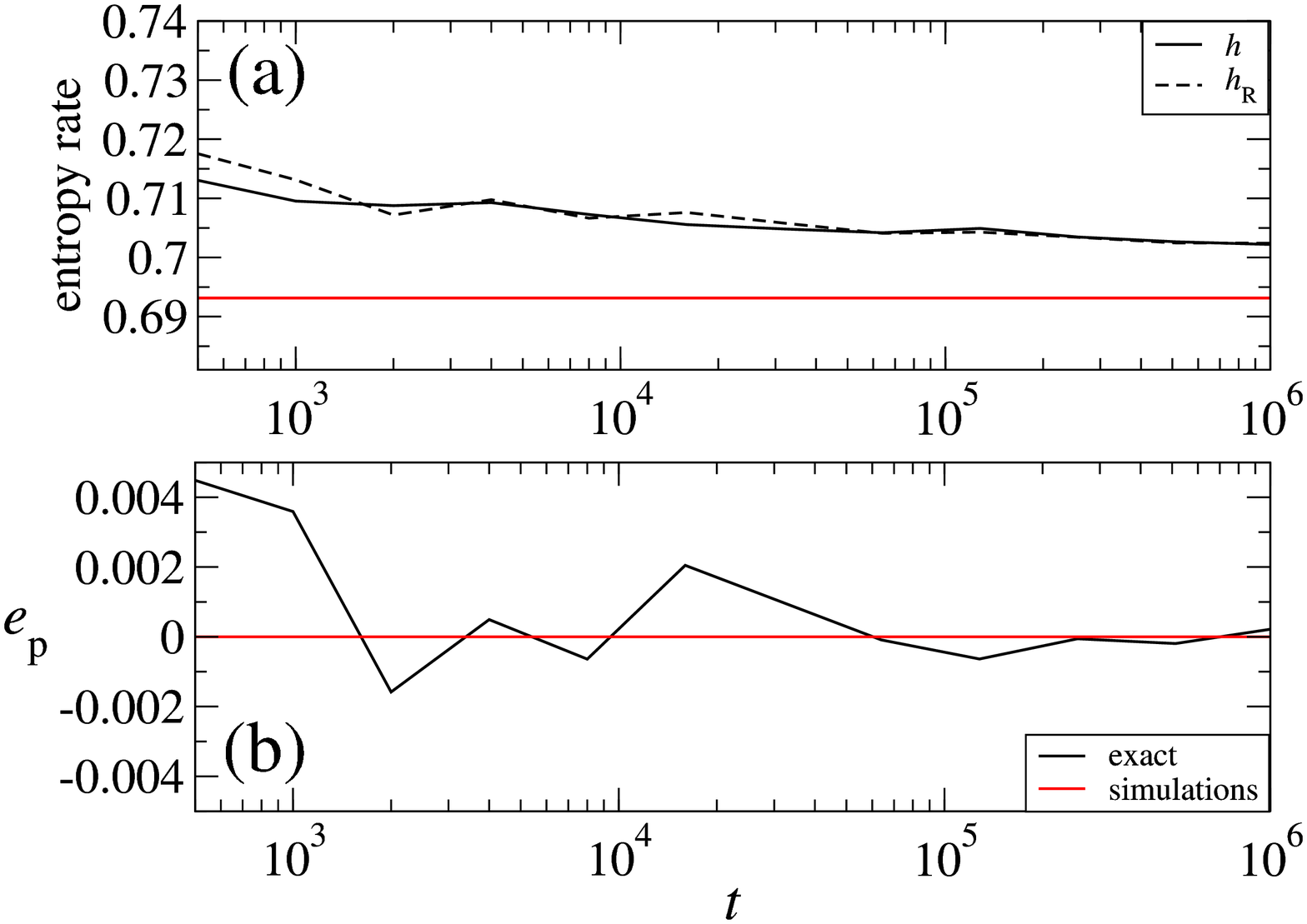}}
\end{center}
     \caption{Estimates of entropy rate and entropy production rate for a reversible Markov chain. (a) Using $m=10^4$ realizations of fixed length $t$ we estimate the entropy rate and the reversed entropy rate using the matching-time estimators given in Eqs.~(\ref{eq:hat-h}) and~(\ref{eq:hat-h}) (solid and dashed lines respectively). The red line stands for the exact value of the entropy rate.  We can observe that the error in the estimation for small $t$ (less that $10^3$ time steps) is of the same order of magnitude as for large $t$ (for $10^6$ time steps). (b) We should notice that the entropy production rate can be accurately predicted for moderately small $t$.  In this case, due to reversibility, the entropy rate and the reversed entropy rate are the same and therefore the entropy production rate is zero. 
     }
\label{fig:fig01}
\end{figure}
%

The first test we implement consists in obtaining $m$ realizations of the process of fixed (non random) length $t$. This numerical experiment is designed to test how accurate is the convergence of the entropy rate through the matching time estimator without considering the randomness in the length of the time series. For the first numerical experiment we simulate the above defined Markov chain for $p = 0.5$. In this case we expect that the entropy rate and the reversed entropy rate be equal, thus obtaining a vanishing entropy production rate. Then we simulate trajectories (time-series) of several lengths, ranging from $t = 500$ time steps up to $t = 10^6$ time steps. For each fixed length $t$ we obtain $m=10^4$ different realizations of the process and for every realization we compute the corresponding matching-time and reversed matching-time. This procedure gives us a sample set of $m=10^4$ realizations of matching-times and a set of $m=10^4$ realizations  of reversed matching-times. These sample sets are then used to obtain the corresponding estimations of the entropy rate and the reversed entropy rate. In Fig.~\ref{fig:fig01}a we show the behavior of the estimated entropy rate and reversed entropy rate as a function of the length of the series. We can observe that the error for short sequences (i.e., for $ t \sim 10^3$ or below) is approximately of the same order of magnitude as for large sequences ($t\sim 10^6$).This means that the matching-time estimator for entropy rate has some stability in its accuracy when the length of the time series changes up to three order of magnitude, making it feasible the entropy estimations for time  series of random length varying from moderately short lengths (around $10^3$ time steps)  up to large lengths without worrying about the introduction of large errors for considering small length time series. This effect is more evident when looking at the estimation of the entropy production rate. For $p=0.5$ the Markov chain is reversible and the corresponding entropy production rate is zero. In Fig.~\ref{fig:fig01}b we display the behavior of the estimated entropy production rate as a function of the size of the series $t$. We should notice that in this case the convergence is fast, since the estimated entropy production rate goes to zero (in the average) for time series lengths of the order of $10^3$ time steps. This property makes the matching-time estimator adequate to test the irreversibility of a series even for moderately short length time series.

\begin{figure}[t]
\begin{center}
\scalebox{0.35}{\includegraphics{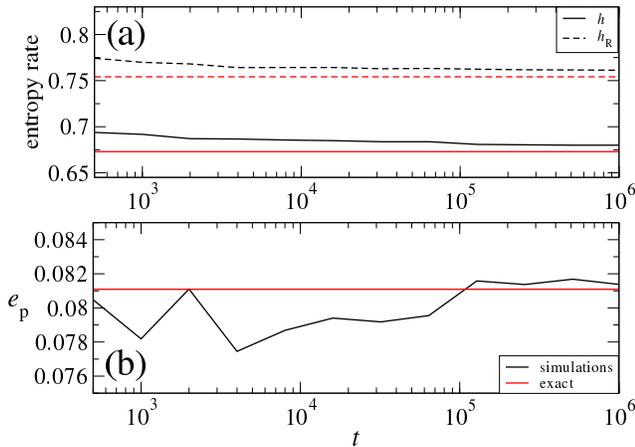}}
\end{center}
     \caption{Estimates of entropy rate and entropy production rate for an irreversible Markov chain. (a) Using $m=10^4$ realizations of fixed length $t$ we estimate the entropy rate and the reversed entropy rate using the matching-time estimators given in Eqs.~(\ref{eq:hat-h}) and~(\ref{eq:hat-h}) (solid and dashed lines respectively). The solid red line and the dashed red lines stands for the exact value of the entropy rate and the reversed entropy rated respectively.  We can observe that the error in both of these estimations for small $t$ (less that $10^3$ time steps) is of the same order of magnitude as for large $t$ (for $10^6$ time steps). (b) We should notice that the estimated entropy production rate (solid line) can be accurately predicted for moderately small $t$. In this case the exact value of entropy production rate is denoted by the solid red line, which is the difference between $h$ and $h_\mathrm{R}$.
     }
\label{fig:fig02}
\end{figure}
%

The second test we implement consists in estimating the entropy rate and reversed entropy rate for time series coming from an irreversible Markov chain. For this case we use the three-states Markov chain model introduced above for $p=0.60$. As in the case $p=0.50$ we perform the estimation of entropy rate without considering the randomness in the length of the time series.  We simulate trajectories of several lengths, ranging from $t = 500$ to $t = 10^6$ time steps. For each fixed length $t$ we obtain $m=10^4$ different realizations of the process and for every realization we compute the corresponding matching-time and reversed matching-time. These sample sets are then used to obtain the corresponding estimations of the entropy rate and the reversed entropy rate through the matching-times as described in Sec.~\ref{ssec:estimation}. In Fig.~\ref{fig:fig02}a we show the behavior of the estimated entropy rate and reversed entropy rate as a function of the length of the series. As in the case $p=0.50$, for the case $p=0.60$ we see that the error for short sequences (for $ t \sim 10^3$) is approximately of the same order of magnitude as for large sequences ($t\sim 10^6$), implying that the accuracy of the estimations is stable a we move from short to large lengths. In Fig.~\ref{fig:fig02}b we show the behavior of the estimated entropy production rate as a function of the series length $t$. We can see that the accuracy of $\hat{e}_\mathrm{p}$ also maintains within the same order of magnitude as we increase the series length from moderately short ($t \sim 10^3$) to large ($t \sim 10^6$). According to these numerical test, it seem that the entropy rate estimator based on matching-times are adequate for testing irreversibility for random length time-series.

\begin{figure}[t]
\begin{center}
\scalebox{0.35}{\includegraphics{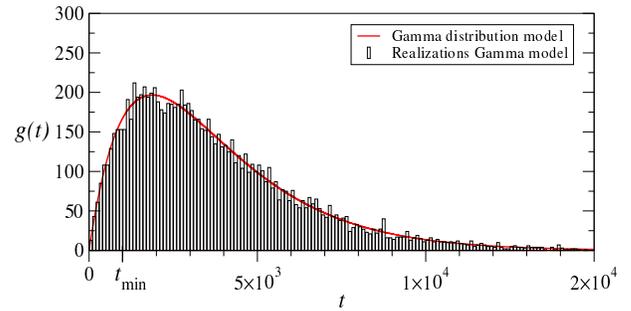}}
\end{center}
     \caption{Random length distribution. We show the histogram of $10^4$ realizations of random lengths (black bars) using the Gamma distribution model (solid red line) given in Eqs.~(\ref{eq:gamma-model-G}) and~(\ref{eq:gamma-model}). The parameters are chosen as $k=1$ and $\lambda = 1/1921$. The latter corresponds to the inverse mean length of the coding sequences of the human genome which where analyzed to test irreversibility. This choice was done to mimic the statistical properties of the real random lengths in order to implement control test with artificial sequences. Notice that the histogram starts at $t_\mathrm{min} = 10^3$. This is because we ignore all sequences below $t_\mathrm{min}$ for performing the entropy estimations.
     }
\label{fig:fig03}
\end{figure}
%

\subsection{\label{ssec:random-length}{Entropy estimation tests for random length time series}}

The next numerical experiment is designed to test the estimators for $h$ and $h_\mathrm{R}$ given in Eqs.~(\ref{eq:hat-h}) and~(\ref{eq:hat-h}), but now considering random length time series. To this end we propose a model distribution to generate the random length time series of the three-states Markov chain. The model we use to generate random lengths is a discrete Gamma distribution~\cite{chakraborty2012discrete} whose probability function $g(t)$, for $t\in \mathbb{N}$ is defined as
\begin{equation}
\label{eq:gamma-model}
g(t) = G(t) - G(t-1),
\end{equation}
where $G(x)$ is the (cumulative) distribution function of a (continuous) random variable with Gamma distribution,
\begin{equation}
\label{eq:gamma-model-G}
G(x) := \int_0^x \frac{\lambda \left( \lambda x\right)^k e^{-\lambda x}}{\Gamma(k)}
\end{equation}
This model was chosen because it reproduce  the main statistical features of the length distribution of real coding sequences  of human genome (see Sec.~\ref{sec:testing-DNA} below). Based on observations of real genomic sequences we choose the parameter $k=1$ and $\lambda = 1/1921$, the latter being the inverse of the mean length coding sequences of human genome.  In Fig.~\ref{fig:fig03} we show the histogram of $m=10^4$ realization of random length using this model. As we will see below, this histogram mimics the main qualitative behavior of the histogram of the random lengths coming from coding sequences of the human DNA. 

Next we generate the random length time series. First we generate a random length $t$ using the Gamma model distribution. Then, we start to generate the time series by simulating the Markov chain and stopping the process after $t$ time steps. It is worth mentioning that we implemented an additional condition on $t$. Actually we limited the value of $t$ to lie in the interval $10^3 \leq t \leq 1.5 \times 10^4$. This condition is imposed to avoid $t$ values too short, in order to have a better control of the error, which is according to the tests implemented in Sec.~\ref{ssec:fixed-length}.

\begin{figure}[t]
\begin{center}
\scalebox{0.5}{\includegraphics{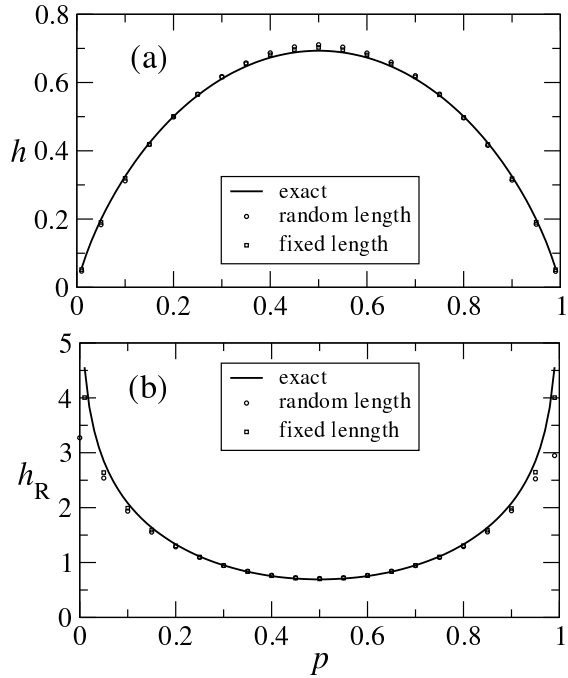}}
\end{center}
     \caption{Estimations of entropy rates and reversed entropy rate for a Markov chain. (a) We show the exact entropy rate $h$ (solid line) as a function of the parameter $p$ of the Markov chain defined in Sec.~\ref{ssec:markov}. We estimate $h$  (open circles) using the method of matching-times for random lengths introduced  in Sec.~\ref{ssec:fixed-length} using the Gamma model for generating the random lengths. We also show the corresponding estimations of entropy rate but using fixed length time series (open squares) for a length of $t=10^6$ time steps (b) The same as in panel (a) for the reversed entropy rate. We should notice that both estimations, fixed and random length time series, gives estimations consistent with the exact corresponding values.
     }
\label{fig:fig04}
\end{figure}
%

We then use this procedure to generate $m=10^4$ time series using the above described Markov chain for several values of the parameter $p$, ranging from $p=0.01$ to $p=0.99$. Next, for every fixed  $p$ we use the entropy estimators~(\ref{eq:hat-h}) and~(\ref{eq:hat-h}) to estimate $h$ and $h_\mathrm{R}$. In Fig~\ref{fig:fig04} we show the estimations of the entropy rate (Fig.~\ref{fig:fig04}a) and the reversed entropy rate (Fig.~\ref{fig:fig04}b) using this procedure (open circles) compared with their exact counterparts (solid black lines) using the corresponding  formulas given in Eqs.~(\ref{eq:h-ex}) and~(\ref{eq:hr-ex}). In order to analyze the performance of these estimations based on random length time series we also display entropy rate (open squares in Fig.~\ref{fig:fig04}a) and the reversed entropy rate (open squares in Fig.~\ref{fig:fig04}b) estimations using fixed-length time series as it was done in Sec.~\ref{ssec:fixed-length}. The length used to obtain the corresponding entropy estimations was $t =  10^6$ time steps.  It is important to stress that for random-length estimations  we used time series whose length lie in the interval  $10^3 \leq t \leq 1.5\times 10^4$). Notice that despite the large difference in the lengths of the time series used for the estimations (a difference of around  two orders of magnitude), we have that the accuracy of the random-length estimations is comparable to the one of the fixed-length estimations. Moreover, as we can appreciate from Fig.~\ref{fig:fig04}, these estimations are both comparable with the corresponding  exact entropy rates. This allows us to state that the proposed estimators based on matching-times seem to be adequate to implement it in real scenarios in which the time series are moderately short and random.

\section{\label{sec:testing-DNA}Testing irreversibility of DNA sequences}


Next we turn out to the problem of determining if the coding sequences of human DNA is reversible or not and if this irreversibility (if any) is comparable to the corresponding degree of irreversibility of the whole human genome. It is important to remark the statistical properties of DNA has been previously studied from several points of view~\cite{rsg2016symbolic}, and in particular the irreversible character of human genome has been analyzed in Refs.~\onlinecite{Provata2014,salgado2021estimating}. It is clear that understanding this characteristic might give us a deeper understanding of the physics of DNA, such as protein diffusion along the DNA~\cite{Gorman2008Visualizing,Mirny2009How,rsg2019noise,cocho2003replication} among other phenomena.

\begin{figure}[t]
\begin{center}
\scalebox{0.45}{\includegraphics{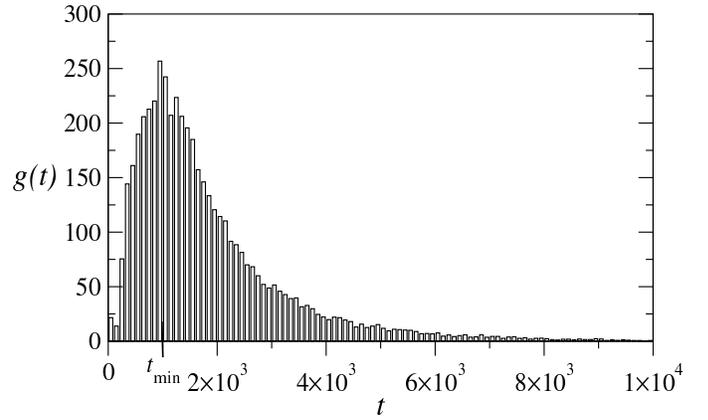}}
\end{center}
     \caption{Random length distribution for coding sequences of \emph{Homo sapiens}. We show the frecuency histogram as a function of the length of the coding sequences. We should notice that the most frequent length we found in coding sequences is nearly $10^3$ bp. Actually, the mean length of the coding sequences is $1970$ bp. For the entropy estimation analysis we only take into account sequences larger than $t_\mathrm{min} = 10^3$ bp. Shorter sequences are excluded to fulfill as most as possible the validity of the central limit theorem.
          }
\label{fig:fig05}
\end{figure}
%

In Refs.\onlinecite{Provata2014,salgado2021estimating} it was found that the entropy production rate estimated from human genome (including coding and non-coding sequences) was around $e_\mathrm{p} \approx 0.07$, a value which is significantly non-zero.  These studies reveal that real genomic sequences might be indeed irreversible, however still remains open the question if the coding part of the genome has a degree of irreversibility larger or lower that the whole genome. In order to shed some light on this question we analyzed the irreversibility of the coding part of the human genome with the technique introduced in this work. To do this, we obtained the coding sequences of all the chromosomes of \emph{Homo sapiens} from the  GenBank database~\cite{benson1997genbank}.

\begin{figure}[t]
\begin{center}
\scalebox{0.35}{\includegraphics{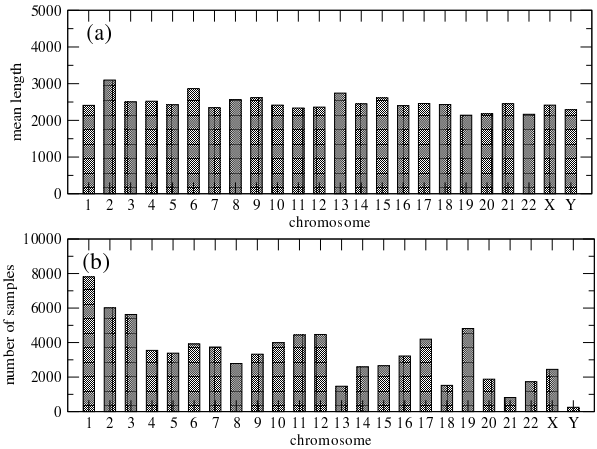}}
\end{center}
     \caption{
     Mean length and number of sample sequences for\emph{Homo sapiens}. (a) We show the mean length of the coding sequences extracted from every chromosome.For the analysis we took into account only coding sequences larger than $10^3$ bp. (b) Number of coding sequences contained in every chromosome. We notice that there are chromosomes with a low number of coding sequences such as the chromosomes $13$, $21$ and $Y$. This fact might introduce statistical errors in estimations. Despite this fact, the entropy estimations even for these chromosomes gives estimations consistent with the estimations from the rest of chromosomes. 
               }
\label{fig:fig06}
\end{figure}

First we should stress the fact that the coding sequences of human genome have different lengths which can be considered as random. The corresponding empiric distribution can be appreciated in Fig.~\ref{fig:fig05} where we display the frequency histogram as a function of the length of the sequence in base-pairs (bp). It is clear that the most frequent length we found in the sample set of coding sequences is of the order of $t\sim 10^3$ bp; actually, the mean length $\bar{t}$ of the coding sequences is $\bar{t} = 1970$ bp. For our analysis we discard the sequences whose length is lower than $t_\mathrm{min} = 10^3$ bp . The latter was done to avoid errors due to short sequences since we need that the sequences be large enough to satisfy the central limit theorem. Taking into account only sequences larger than $ 10^3$ bp coming from all the chromosomes, we analyzed $80671$ coding sequences with a mean length $3360$ bp. In Fig.~\ref{fig:fig06}a we show the mean length  $t$ of the coding sequences from every chromosome. In Fig.~\ref{fig:fig06}b we observe the number of coding sequences contained in every chromosome. We can see that there are chromosomes with a low number of coding sequences such as the chromosomes $13$, $21$ and $Y$. This fact might introduce statistical errors in estimations. Despite this fact, as we will see below, the entropy estimations even for these chromosomes gives estimations consistent with the estimations from the rest of chromosomes.

%

\begin{figure}[t]
\begin{center}
\scalebox{0.35}{\includegraphics{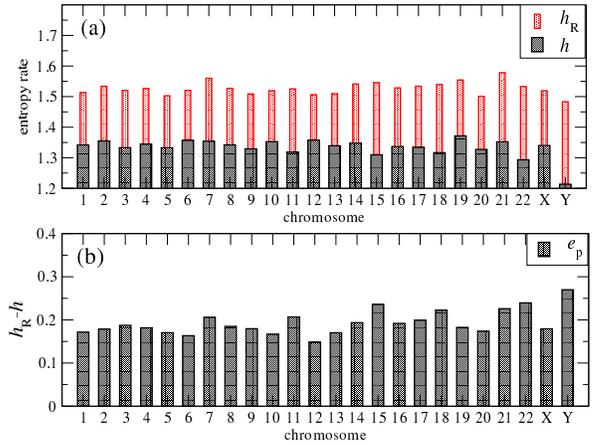}}
\end{center}
     \caption{Entropy estimations for coding sequences of \emph{Homo sapiens}. (a) We show the estimations of the entropy rate (black bars) and the reversed entropy rate (red bars) for the coding part of every chromosome. We should notice that the reversed entropy rate is larger than the entropy rate, which means that the coding sequences are spatially irreversible. (b) We show the behavior of the estimated entropy production rate for every chromosome. We can observe that the value of $e_\mathrm{p}$ deviates moderately on the chromosomes, thus allowing us to say that the mean value of the entropy production rate that we estimate is representative of the coding part of the human genome. 
          }
\label{fig:fig07}
\end{figure}

Next we proceed to estimate the entropy rate and the reversed entropy rate for every chromosome separately. In Fig.~\ref{fig:fig07}a we can see the estimations of $h$ (black bars) and $h_\mathrm{R}$  (red bars) for every human chromosome. We should see that the reversed entropy rate is significantly larger than the entropy rate thus allowing us to see that there is certain irreversibility of the coding sequences. In Fig.~\ref{fig:fig07}b we display the estimated entropy production rate ($e_\mathrm{p} = h_\mathrm{R}-h$) for every chromosome. We can see that the entropy production rate is around $0.2$ and actually the mean $\hat e_\mathrm{p} $ we obtain from these data is $e_\mathrm{p} \approx 0.1928 $ In Table~\ref{tab:tab1} we resume the values of the estimations of entropy rate and entropy production rate for both coding and non-coding sequences of \emph{Homo sapiens}. The values for non-coding sequences was obtained from Refs.~\onlinecite{Provata2014,salgado2021estimating}.

We clearly see that the estimated entropy production rate for coding sequences that we obtained here is larger than twice the entropy production rate reported for the analyzed sequences of \emph{Homo sapiens} which include both coding and non-coding. Since the non-coding sequences in \emph{Homo sapiens} is around the $98\%$ of the whole genome we can say that the estimations reported in Refs.~\onlinecite{Provata2014,salgado2021estimating} are representative of the non-coding genomic material. Therefore, our results allow to state that the degree of irreversibility of coding sequences is significantly larger than the non-coding material. This would mean that the non-coding part of the genome might possibly underwent several processes that lead to lose its irreversibility, such as random mutations or any other process (or thermodynamic force) that drives the non-coding genetic material to the equilibrium in the sense of reversibility.

%

%
%
\begin{table}
\caption{\label{tab:tab1}Entropy rate and entropy production rate for coding and non-coding DNA of \emph{Homo sapiens}. }
\begin{ruledtabular}
\begin{tabular}{ccc}
                 &  coding & non-coding   \\ 
$\hat h$              &   $1.3334$   &  $1.277$\footnotemark[1] \\
$\hat h_\mathrm{R}$   &   $1.5262$   &  $1.352$\footnotemark[1]   \\
$\hat e_\mathrm{p}$   &   $0.1928$   &  $0.075$\footnotemark[2]   \\
$\mbox{STD}(e_\mathrm{p})/\hat e_\mathrm{p}$  &   $0.1442$\footnotemark[3]   &  $-$   \\
\end{tabular}
\end{ruledtabular}
\footnotetext[1]{From Ref.~\onlinecite{salgado2021estimating} }
\footnotetext[2]{From Ref.~\onlinecite{Provata2014}. The corresponding value reported in Ref.~\onlinecite{salgado2021estimating} is $e_\mathrm{p} \approx 0.077$. }
\footnotetext[3]{The standard deviation of $e_\mathrm{p}$ is computed with respect to the estimated values from the chromosomes. It represents the mean deviation of the entropy production rate from one chromosome to another.}
\end{table}
%
%

\bigskip

\section{Conclusions}
\label{sec:conclusions}

We have introduced the maximum likelihood estimator of entropy rate based on the recurrence properties of the system. In particular we made use of the so-called matching times which gives an estimator for the inverse entropy rate. According to the central limit theorem~\cite{Kon}, this estimator converges to the normal distribution if it is appropriately normalized. We used this fact to obtain a maximum likelihood estimator when we have time series of several size i.e., when we have a sample set of time series of random duration. We showed through numerical simulations that the proposed estimator gives accurate enough estimations of entropy rate and reversed entropy rate in the case of Markov chain even for moderately short time series (of around $t^3$ time steps). Once we tested the estimator using time series of random length obtained from both, a reversible and an irreversible Markov chain, we proceed to implement our method to determine the degree of irreversibility of coding sequences of human genome. We observed that the coding sequences of \emph{Homo sapiens} have an entropy production rate  $\hat e_\mathrm{p}  = 0.1928$. Interestingly we saw that this value for $\hat e_\mathrm{p}$ does not deviates too much from one chromosome to another, i.e., the degree of irreversibility of all the coding part of all the chromosomes of the human genome is approximately the same. This fact suggest that the irreversibility character of the coding sequences might be result of a evolutive process that lead the coding sequences to its current degree of irreversibility. Moreover, other studies revealed that the non-coding part of the human genome has a degree of irreversibility that is lower than the one we report here for the coding sequences. This would mean that the non-coding part of the genome, might underwent some process that lead it to lower its degree of irreversibility in the course of evolution, i.e., the non-coding part of the genome might possibly be under the influence of some thermodynamic force that drives the sequences to the equilibrium in the sense of irreversibility. Clearly, understanding the origin of irreversibility of real genomes requiere a much more deeper statistical analysis from the point of view of genome evolution.

\bigskip

\begin{acknowledgments}
This work was supported by CONACYT through grant FORDECYT-PRONACES/1327701/2020.
\end{acknowledgments}

\appendix

\section{Maximum Likelihood estimator for inhomogeneous samples.}
\label{ape:ape1}

Let us start by assuming that we have a normal random variable $X$ whose distribution depends on a value $t$. We assume this parameter $t$ is a realization of a random variable $T$ whose distribution is denoted by $g(t,\theta)$, where $\theta$ is a (unknown) parameter to be estimated. In terms of $t$, the mean and variance of $X$ are given by 
\begin{eqnarray}
\mathbb{E}[X] &=& \frac{1}{h},
\\
\mbox{Var}(X) &=& \frac{\sigma^2 }{h^{3}\log(t)},
\end{eqnarray}
and the corresponding probability density function of $X$ can be written as,
\begin{equation}
f(x;h,\sigma,t) := \frac{ h^{3}\log(t)}{\sqrt{2\pi \sigma^2 }}\exp\left(- \frac{\left( x - \frac{1}{h}\right)^2  h^{3}\log(t)}{2\sigma^2}\right).
\end{equation}

Since the parameter $t$ is a realization of the random variable $T$, it is clear that $f(x;h,\sigma,t)$ is the conditioned distribution of $X$ given $T$. Therefore, assuming that $T$ is independent, it is clear that the joint probability density function $f_{X,Y}(x,y)$ of $(X,T)$ is given by
\begin{equation}
f_{X,Y}(x,y) = f(x;h,\sigma,t)g(t,\theta).
\end{equation}
where the variables $h$, $\sigma^2$, and $\theta$ are parameters to be estimated from a given sample.

Next we consider a sample set $\mathcal{S}$ of realizations of $(X,T)$ as follows,
\begin{equation}
\mathcal{S} := \{ (x_i,t_i) : 1\leq i \leq m\},
\end{equation}
which will be used to estimate  $h$, $\sigma^2$, and $\theta$ by means of the maximum likelihood method.

Once we have stated the joint density function of $(X,T)$, the likelihood function for the sample set $\mathcal{S}$ can be written as follows,
\begin{equation}
L(h,\sigma^2; \mathcal{S}) =\prod_{i=1}^m  f(x_i;h,\sigma,t_i)g(t_i,\theta)\Delta x. 
\end{equation}
The log-likelihood function is therefore given by
\begin{eqnarray}
\log L &=& \sum_{i=1}^m \bigg[ - \frac{\left( x_i - \frac{1}{h}\right)^2 h^3\log(t_i) }{2\sigma^2} 
+\frac{1}{2} \log \left( \frac{h^3\log(t_i)}{ \sigma^2} \right)
\nonumber
\\
&+& \log \left( g(t_i,\theta) \right) + \log\left( \frac{\Delta x }{\sqrt{2\pi}}\right)  \bigg]. 
\end{eqnarray}

After some calculations it is easy to see that the first derivative of $\log L$ with respect to $h$, $\sigma^2$ and $\theta$ can be expressed as,
\begin{eqnarray}
\frac{\partial \log L}{\partial h} &=& -\frac{3}{2} \sum_{i=1}^m  \frac{\left( x_i - \frac{1}{h}\right)^2 h^2 \log(t_i)}{\sigma^2} -  \sum_{i=1}^m  \frac{\left( x_i - \frac{1}{h}\right) h\log(t_i) }{\sigma^2} 
\nonumber 
\\
&+& \frac{3}{2}\frac{m}{h}, \quad
\\
\frac{\partial \log L}{\partial \sigma^2} &=&  \sum_{i=1}^m  \frac{\left( x_i - \frac{1}{h}\right)^2 h^3 \log(t_i) }{2\sigma^4} - \frac{m}{2 \sigma^2},
\\
\frac{\partial \log L}{\partial \theta} &=& \sum_{i=1}^m \frac{\partial \log g(t_i;\theta) }{\partial \theta}.
\end{eqnarray}

Now, to maximize the log-likelihood function we equate to zero the above partial derivatives. Solving these equations will give us the maximum likelihood estimations for every parameter. Then we obtain,
\begin{eqnarray}
\frac{m \sigma^2 }{h} &=& \sum_{i=1}^m   \left( x_i - \frac{1}{h}\right)^2 h^2 \log(t_i) + \frac{2}{3}\sum_{i=1}^m  \left( x_i - \frac{1}{h}\right) h \log(t_i),
\nonumber 
\\
\label{eq:to-solve-1}
& &
\\
m \sigma^2 &=& \sum_{i=1}^m  \left( x_i - \frac{1}{h}\right)^2 h^3\log(t_i)  , \qquad
\label{eq:to-solve-2}
\\
0 &=& \sum_{i=1}^m \frac{\partial \log g(t_i;\theta) }{\partial \theta} . \qquad
\label{eq:to-solve-3}
\end{eqnarray}

Notice that the last equation involves uniquely the parameter $\theta$. This means that the maximum likelihood estimation for $\theta$ can be obtained separately from the problem of estimating $h$ and $\sigma^2$. This was actually expected due to the fact that $T$ is an independent random variable and therefore the parameters of the model can be estimated without the information about the realizations $x_i$. The latter means that the distribution of random duration of the time series can be estimated independently of the estimations of any other property of the series, which in the present case is entropy rate and the entropy production rate of the process.

For the sake of clearness we introduce the following short-hand notations 
\begin{eqnarray}
\hat{a} := \frac{1}{m} \sum_{i=1}^m x_i \log(t_i),
\\
\hat{b} := \frac{1}{m} \sum_{i=1}^m \log(t_i),
\\
\label{eq:ape:xi-def}
\hat{\xi} := \frac{1}{m} \sum_{i=1}^m \log(t_i) \left( x_i - \frac{1}{h}\right)^2.
\end{eqnarray}
In terms of $\hat{a}$, $\hat{b}$, and  $\hat{\xi}$ we can rewrite Eqs.~(\ref{eq:to-solve-1}) and~(\ref{eq:to-solve-2}) as
\begin{eqnarray}
\frac{\sigma^2 }{h} &=&  h^2 \hat{\xi} + \frac{2}{3} \left(\hat{a} h - \hat{b}\right),
\label{eq:to-solve-int1}
\\
\sigma^2 &=& h^3\hat{\xi}  , \qquad
\label{eq:to-solve-int2}
\end{eqnarray}
Next, from Eq.~(\ref{eq:to-solve-int2}) we see that $\hat{\xi} = \sigma^2/h^3$, which can be substituted into Eq.~(\ref{eq:to-solve-int1}) to obtain,
\begin{equation}
\frac{\sigma^2 }{h} = \frac{\sigma^2}{h} + \frac{2}{3} \left(\hat{a} h - \hat{b}\right),
\end{equation}
which implies that $\hat{a} h - \hat{b} = 0$, thus obtaining the maximum likelihood estimation $\hat h$ for the parameter $h$,
\begin{equation}
\hat{h} := \frac{ \hat{b}}{ \hat{a}} = \frac{  \frac{1}{m} \sum_{i=1}^m \log(t_i)  }{  \frac{1}{m} \sum_{i=1}^m x_i\log(t_i)}.
\end{equation}
Recalling that $x_i$ is actually defined as $x_i = \ell_i/\log(t_i)$, we can see that $\hat h$ can alternatively written as
\begin{equation}
\hat{h} =  \frac{  \frac{1}{m} \sum_{i=1}^m \log(t_i)  }{  \frac{1}{m} \sum_{i=1}^m \ell_i },
\end{equation}
which was the formula anticipated in Eqs.~(\ref{eq:hat-h}) and~(\ref{eq:hat-hR}).

Now, once we have an estimation for $h$, we can obtain an estimation for $\sigma^2$ through the expression~(\ref{eq:to-solve-int2}). In order to obtain a simplified formula for this parameter, let us perform some calculation. Firs notice that $\hat\xi$ defined in  eq.~(\ref{eq:ape:xi-def}) can be written as,
\begin{eqnarray}
\hat{\xi} &=& \frac{1}{m} \sum_{i=1}^m \log(t_i) \left( x_i^2 - 2\frac{x_i}{h} + \frac{1}{h^2}\right)
\nonumber
\\
&=& \frac{1}{m} \sum_{i=1}^m x_i^2 \log(t_i) - \frac{2}{h} \frac{1}{m} \sum_{i=1}^m x_i \log(t_i) 
+  \frac{1}{h^2} \frac{1}{m} \sum_{i=1}^m  \log(t_i)
\nonumber
\\
&=& \hat{c} - \frac{2\hat{a}^2}{h} +  \frac{ \hat{b}}{h^2},
\end{eqnarray}
where we defined $\hat{c}$ as
\begin{eqnarray}
\hat{c} &:=&   \frac{1}{m} \sum_{i=1}^m x_i^2 \log(t_i).
\end{eqnarray}
Thus, in terms of the sample functions $\hat{a}$, $\hat{b}$, and  $\hat{c}$ we have that an estimation for the parametes $\sigma^2$ can be written as
\begin{equation}
\hat{\sigma}^2 = \hat{h}^3 \hat{c} - 2 \hat{h}^2 \hat{a} +  \hat{h} \hat{b}.
\end{equation}
We can simplify the above expression if we recall that $\hat{h} = \hat{b}/\hat{a}$. Further calculations allows us to write  
\begin{equation}
\hat{\sigma}^2 = \hat{h}^2 \left( \hat{h} \hat{c} -  \hat{a} \right).
\end{equation}

It is clear that the parameter $\sigma $ is associated to the statistical error of $X$ and therefore it would be involved in the error associated to the estimation of $h$. Actually we will not obtain the exact expression of the standard deviation of the estimator $\hat{h}$ but we will made an approximation that will allows us to known the error due to the finiteness of the sample sequences $\{t_i\}$. Since $\sigma/\sqrt{ h^3 \log(t)}$ is an approximation to the error of the inverse entropy rate, it is clear that 
\begin{equation}
\frac{1}{h} \approx \frac{1}{\hat{h}} \pm \frac{\hat{\sigma}}{\sqrt{ \hat{h}^3 \log(t)} }.
\end{equation}
Taking the inverse of the above equality we obtain
\begin{equation}
h \approx \left( \frac{1}{\hat{h}} \pm \frac{\hat{\sigma}}{\sqrt{ \hat{h}^3 \log(t)} }\right)^{-1},
\end{equation}
thus, assuming that $\sigma/\sqrt{ h^3 \log(t)}$ is small, we can expand the right-hand side of the above equation up to first order. Then we obtain
\begin{equation}
h \approx \hat{h}  \pm  \frac{\hat{\sigma}}{\sqrt{ \hat{h} \log(t)} }.
\end{equation}
Finally, since $t$ is a random variable we average the squared error over all the sequence length of the sample, thus obtaining that
\begin{equation}
\hat{\epsilon}^2= \frac{\hat \sigma^2}{{ \hat h } } \frac{1}{m}\sum_{i=1}^m \frac{1}{{\log(t_i)}}.
\end{equation}
This  last expression allows to see that our estimator is consistent with the central limit theorem in the sense that error vanishes if we made $t_i$ tend to infinity. Otherwise, for finite $t$ the error remains finite no matter how many samples we use, i.e., no matter how long is $m$. The latter, as we can see, is a consequence of the fact that there is an intrinsic error due to the finiteness of $t$.


\section*{Data Availability}
The data that supports the findings of this study are available within the article.

\section*{References}
\nocite{*}
\bibliography{Matching_refs}

\begin{thebibliography}{24}%
\makeatletter
\providecommand \@ifxundefined [1]{%
 \@ifx{#1\undefined}
}%
\providecommand \@ifnum [1]{%
 \ifnum #1\expandafter \@firstoftwo
 \else \expandafter \@secondoftwo
 \fi
}%
\providecommand \@ifx [1]{%
 \ifx #1\expandafter \@firstoftwo
 \else \expandafter \@secondoftwo
 \fi
}%
\providecommand \natexlab [1]{#1}%
\providecommand \enquote  [1]{``#1''}%
\providecommand \bibnamefont  [1]{#1}%
\providecommand \bibfnamefont [1]{#1}%
\providecommand \citenamefont [1]{#1}%
\providecommand \href@noop [0]{\@secondoftwo}%
\providecommand \href [0]{\begingroup \@sanitize@url \@href}%
\providecommand \@href[1]{\@@startlink{#1}\@@href}%
\providecommand \@@href[1]{\endgroup#1\@@endlink}%
\providecommand \@sanitize@url [0]{\catcode `\\12\catcode `\$12\catcode
  `\&12\catcode `\#12\catcode `\^12\catcode `\_12\catcode `\%12\relax}%
\providecommand \@@startlink[1]{}%
\providecommand \@@endlink[0]{}%
\providecommand \url  [0]{\begingroup\@sanitize@url \@url }%
\providecommand \@url [1]{\endgroup\@href {#1}{\urlprefix }}%
\providecommand \urlprefix  [0]{URL }%
\providecommand \Eprint [0]{\href }%
\providecommand \doibase [0]{http://dx.doi.org/}%
\providecommand \selectlanguage [0]{\@gobble}%
\providecommand \bibinfo  [0]{\@secondoftwo}%
\providecommand \bibfield  [0]{\@secondoftwo}%
\providecommand \translation [1]{[#1]}%
\providecommand \BibitemOpen [0]{}%
\providecommand \bibitemStop [0]{}%
\providecommand \bibitemNoStop [0]{.\EOS\space}%
\providecommand \EOS [0]{\spacefactor3000\relax}%
\providecommand \BibitemShut  [1]{\csname bibitem#1\endcsname}%
\let\auto@bib@innerbib\@empty
\bibitem [{\citenamefont {Daw}, \citenamefont {Finney},\ and\ \citenamefont
  {Kennel}(2000)}]{daw2000symbolic}%
  \BibitemOpen
  \bibfield  {author} {\bibinfo {author} {\bibfnamefont {C.}~\bibnamefont
  {Daw}}, \bibinfo {author} {\bibfnamefont {C.}~\bibnamefont {Finney}}, \ and\
  \bibinfo {author} {\bibfnamefont {M.}~\bibnamefont {Kennel}},\ }\bibfield
  {title} {\enquote {\bibinfo {title} {Symbolic approach for measuring temporal
  “irreversibility”},}\ }\href@noop {} {\bibfield  {journal} {\bibinfo
  {journal} {Physical Review E}\ }\textbf {\bibinfo {volume} {62}},\ \bibinfo
  {pages} {1912} (\bibinfo {year} {2000})}\BibitemShut {NoStop}%
\bibitem [{\citenamefont {Latora}\ and\ \citenamefont
  {Baranger}(1999)}]{latora1999kolmogorov}%
  \BibitemOpen
  \bibfield  {author} {\bibinfo {author} {\bibfnamefont {V.}~\bibnamefont
  {Latora}}\ and\ \bibinfo {author} {\bibfnamefont {M.}~\bibnamefont
  {Baranger}},\ }\bibfield  {title} {\enquote {\bibinfo {title}
  {Kolmogorov-sinai entropy rate versus physical entropy},}\ }\href@noop {}
  {\bibfield  {journal} {\bibinfo  {journal} {Phys. Rev. Lett.}\ }\textbf
  {\bibinfo {volume} {82}},\ \bibinfo {pages} {520} (\bibinfo {year}
  {1999})}\BibitemShut {NoStop}%
\bibitem [{\citenamefont {Porporato}, \citenamefont {Rigby},\ and\
  \citenamefont {Daly}(2007)}]{PorpoAl}%
  \BibitemOpen
  \bibfield  {author} {\bibinfo {author} {\bibfnamefont {A.}~\bibnamefont
  {Porporato}}, \bibinfo {author} {\bibfnamefont {J.~R.}\ \bibnamefont
  {Rigby}}, \ and\ \bibinfo {author} {\bibfnamefont {E.}~\bibnamefont {Daly}},\
  }\bibfield  {title} {\enquote {\bibinfo {title} {Irreversibility and
  fluctuation theorem in stationary time series},}\ }\href {\doibase
  10.1103/PhysRevLett.98.094101} {\bibfield  {journal} {\bibinfo  {journal}
  {Phys. Rev. Lett.}\ }\textbf {\bibinfo {volume} {98}},\ \bibinfo {pages}
  {094101} (\bibinfo {year} {2007})}\BibitemShut {NoStop}%
\bibitem [{\citenamefont {Rold\'an}\ and\ \citenamefont
  {Parrondo}(2012)}]{RolP}%
  \BibitemOpen
  \bibfield  {author} {\bibinfo {author} {\bibfnamefont {E.}~\bibnamefont
  {Rold\'an}}\ and\ \bibinfo {author} {\bibfnamefont {J.~M.~R.}\ \bibnamefont
  {Parrondo}},\ }\bibfield  {title} {\enquote {\bibinfo {title} {Entropy
  production and kullback-leibler divergence between stationary trajectories of
  discrete systems},}\ }\href {\doibase 10.1103/PhysRevE.85.031129} {\bibfield
  {journal} {\bibinfo  {journal} {Phys. Rev. E}\ }\textbf {\bibinfo {volume}
  {85}},\ \bibinfo {pages} {031129} (\bibinfo {year} {2012})}\BibitemShut
  {NoStop}%
\bibitem [{\citenamefont {Gaspard}(2004)}]{gaspard2004time}%
  \BibitemOpen
  \bibfield  {author} {\bibinfo {author} {\bibfnamefont {P.}~\bibnamefont
  {Gaspard}},\ }\bibfield  {title} {\enquote {\bibinfo {title} {Time-reversed
  dynamical entropy and irreversibility in markovian random processes},}\
  }\href@noop {} {\bibfield  {journal} {\bibinfo  {journal} {J. of Stat.
  Phys.}\ }\textbf {\bibinfo {volume} {117}},\ \bibinfo {pages} {599--615}
  (\bibinfo {year} {2004})}\BibitemShut {NoStop}%
\bibitem [{\citenamefont {Costa}, \citenamefont {Goldberger},\ and\
  \citenamefont {Peng}(2005)}]{costa2005broken}%
  \BibitemOpen
  \bibfield  {author} {\bibinfo {author} {\bibfnamefont {M.}~\bibnamefont
  {Costa}}, \bibinfo {author} {\bibfnamefont {A.~L.}\ \bibnamefont
  {Goldberger}}, \ and\ \bibinfo {author} {\bibfnamefont {C.-K.}\ \bibnamefont
  {Peng}},\ }\bibfield  {title} {\enquote {\bibinfo {title} {Broken asymmetry
  of the human heartbeat: loss of time irreversibility in aging and disease},}\
  }\href@noop {} {\bibfield  {journal} {\bibinfo  {journal} {Phys. Rev. Lett.}\
  }\textbf {\bibinfo {volume} {95}},\ \bibinfo {pages} {198102} (\bibinfo
  {year} {2005})}\BibitemShut {NoStop}%
\bibitem [{\citenamefont {Salgado-Garcia}\ and\ \citenamefont
  {Maldonado}(2021)}]{salgado2021estimating}%
  \BibitemOpen
  \bibfield  {author} {\bibinfo {author} {\bibfnamefont {R.}~\bibnamefont
  {Salgado-Garcia}}\ and\ \bibinfo {author} {\bibfnamefont {C.}~\bibnamefont
  {Maldonado}},\ }\bibfield  {title} {\enquote {\bibinfo {title} {Estimating
  entropy rate from censored symbolic time series: A test for
  time-irreversibility},}\ }\href@noop {} {\bibfield  {journal} {\bibinfo
  {journal} {Chaos: An Interdisciplinary Journal of Nonlinear Science}\
  }\textbf {\bibinfo {volume} {31}},\ \bibinfo {pages} {013131} (\bibinfo
  {year} {2021})}\BibitemShut {NoStop}%
\bibitem [{\citenamefont {Provata}, \citenamefont {Nicolis},\ and\
  \citenamefont {Nicolis}(2014)}]{Provata2014}%
  \BibitemOpen
  \bibfield  {author} {\bibinfo {author} {\bibfnamefont {A.}~\bibnamefont
  {Provata}}, \bibinfo {author} {\bibfnamefont {C.}~\bibnamefont {Nicolis}}, \
  and\ \bibinfo {author} {\bibfnamefont {G.}~\bibnamefont {Nicolis}},\
  }\bibfield  {title} {\enquote {\bibinfo {title} {{DNA viewed as an
  out-of-equilibrium structure}},}\ }\href {\doibase
  10.1103/PhysRevE.89.052105} {\bibfield  {journal} {\bibinfo  {journal} {Phys.
  Rev. E}\ }\textbf {\bibinfo {volume} {89}},\ \bibinfo {pages} {052105}
  (\bibinfo {year} {2014})}\BibitemShut {NoStop}%
\bibitem [{\citenamefont {Flanagan}\ and\ \citenamefont
  {Lacasa}(2016)}]{flanagan2016irreversibility}%
  \BibitemOpen
  \bibfield  {author} {\bibinfo {author} {\bibfnamefont {R.}~\bibnamefont
  {Flanagan}}\ and\ \bibinfo {author} {\bibfnamefont {L.}~\bibnamefont
  {Lacasa}},\ }\bibfield  {title} {\enquote {\bibinfo {title} {Irreversibility
  of financial time series: a graph-theoretical approach},}\ }\href@noop {}
  {\bibfield  {journal} {\bibinfo  {journal} {Phys. Lett. A}\ }\textbf
  {\bibinfo {volume} {380}},\ \bibinfo {pages} {1689--1697} (\bibinfo {year}
  {2016})}\BibitemShut {NoStop}%
\bibitem [{\citenamefont {Gonz\'alez-Espinoza}, \citenamefont
  {Mart\'{\i}nez-Mekler},\ and\ \citenamefont {Lacasa}(2020)}]{Gustavo}%
  \BibitemOpen
  \bibfield  {author} {\bibinfo {author} {\bibfnamefont {A.}~\bibnamefont
  {Gonz\'alez-Espinoza}}, \bibinfo {author} {\bibfnamefont {G.}~\bibnamefont
  {Mart\'{\i}nez-Mekler}}, \ and\ \bibinfo {author} {\bibfnamefont
  {L.}~\bibnamefont {Lacasa}},\ }\bibfield  {title} {\enquote {\bibinfo {title}
  {Arrow of time across five centuries of classical music},}\ }\href {\doibase
  10.1103/PhysRevResearch.2.033166} {\bibfield  {journal} {\bibinfo  {journal}
  {Phys. Rev. Research}\ }\textbf {\bibinfo {volume} {2}},\ \bibinfo {pages}
  {033166} (\bibinfo {year} {2020})}\BibitemShut {NoStop}%
\bibitem [{\citenamefont {Li}\ and\ \citenamefont
  {Graur}(1991)}]{li1991fundamentals}%
  \BibitemOpen
  \bibfield  {author} {\bibinfo {author} {\bibfnamefont {W.-H.}\ \bibnamefont
  {Li}}\ and\ \bibinfo {author} {\bibfnamefont {D.}~\bibnamefont {Graur}},\
  }\href@noop {} {\emph {\bibinfo {title} {Fundamentals of molecular
  evolution}}},\ \bibinfo {number} {576.5 L5}\ (\bibinfo {year}
  {1991})\BibitemShut {NoStop}%
\bibitem [{\citenamefont {Kontoyiannis}(1998)}]{Kon}%
  \BibitemOpen
  \bibfield  {author} {\bibinfo {author} {\bibfnamefont {I.}~\bibnamefont
  {Kontoyiannis}},\ }\bibfield  {title} {\enquote {\bibinfo {title} {Asymptotic
  recurrence and waiting times for stationary processes},}\ }\href {\doibase
  10.1023/A:1022610816550} {\bibfield  {journal} {\bibinfo  {journal} {J.
  Theor. Prob.}\ }\textbf {\bibinfo {volume} {11}},\ \bibinfo {pages}
  {795--811} (\bibinfo {year} {1998})}\BibitemShut {NoStop}%
\bibitem [{\citenamefont {Chazottes}\ and\ \citenamefont {Redig}(2005)}]{ChR}%
  \BibitemOpen
  \bibfield  {author} {\bibinfo {author} {\bibfnamefont {J.-R.}\ \bibnamefont
  {Chazottes}}\ and\ \bibinfo {author} {\bibfnamefont {F.}~\bibnamefont
  {Redig}},\ }\bibfield  {title} {\enquote {\bibinfo {title} {Testing the
  irreversibility of a {G}ibbsian process via hitting and return times},}\
  }\href {\doibase 10.1088/0951-7715/18/6/004} {\bibfield  {journal} {\bibinfo
  {journal} {Nonlinearity}\ }\textbf {\bibinfo {volume} {18}},\ \bibinfo
  {pages} {2477--2489} (\bibinfo {year} {2005})}\BibitemShut {NoStop}%
\bibitem [{\citenamefont {Chazottes}\ and\ \citenamefont {Ugalde}(2005)}]{ChU}%
  \BibitemOpen
  \bibfield  {author} {\bibinfo {author} {\bibfnamefont {J.-R.}\ \bibnamefont
  {Chazottes}}\ and\ \bibinfo {author} {\bibfnamefont {E.}~\bibnamefont
  {Ugalde}},\ }\bibfield  {title} {\enquote {\bibinfo {title} {Entropy
  estimation and fluctuations of hitting and recurrence times for {G}ibbsian
  sources},}\ }\href {\doibase 10.3934/dcdsb.2005.5.565} {\bibfield  {journal}
  {\bibinfo  {journal} {Discrete Continuous Dynamical Systems Ser. B}\ }\textbf
  {\bibinfo {volume} {5}},\ \bibinfo {pages} {565--586} (\bibinfo {year}
  {2005})}\BibitemShut {NoStop}%
\bibitem [{\citenamefont {Maldonado}(2015)}]{cesar2015fluctuations}%
  \BibitemOpen
  \bibfield  {author} {\bibinfo {author} {\bibfnamefont {C.}~\bibnamefont
  {Maldonado}},\ }\bibfield  {title} {\enquote {\bibinfo {title} {Fluctuation
  bounds for entropy production estimators in gibbs measures},}\ }\href@noop {}
  {\bibfield  {journal} {\bibinfo  {journal} {J. of Phys. A: Math. and Theor.}\
  }\textbf {\bibinfo {volume} {48}},\ \bibinfo {pages} {045003} (\bibinfo
  {year} {2015})}\BibitemShut {NoStop}%
\bibitem [{\citenamefont {Maes}(1999)}]{Mae}%
  \BibitemOpen
  \bibfield  {author} {\bibinfo {author} {\bibfnamefont {C.}~\bibnamefont
  {Maes}},\ }\bibfield  {title} {\enquote {\bibinfo {title} {The fluctuation
  theorem as a {G}ibbs property},}\ }\href {\doibase 10.1023/A:1004541830999}
  {\bibfield  {journal} {\bibinfo  {journal} {J. Stat. Phys.}\ }\textbf
  {\bibinfo {volume} {95}},\ \bibinfo {pages} {367--392} (\bibinfo {year}
  {1999})}\BibitemShut {NoStop}%
\bibitem [{\citenamefont {Jiang}, \citenamefont {Qian},\ and\ \citenamefont
  {Qian}(2004)}]{Jiang}%
  \BibitemOpen
  \bibfield  {author} {\bibinfo {author} {\bibfnamefont {D.-Q.}\ \bibnamefont
  {Jiang}}, \bibinfo {author} {\bibfnamefont {M.}~\bibnamefont {Qian}}, \ and\
  \bibinfo {author} {\bibfnamefont {M.-P.}\ \bibnamefont {Qian}},\ }\href@noop
  {} {\emph {\bibinfo {title} {Mathematical Theory of Nonequilibrium Steady
  States}}}\ (\bibinfo  {publisher} {Springer},\ \bibinfo {year}
  {2004})\BibitemShut {NoStop}%
\bibitem [{\citenamefont {Chakraborty}\ and\ \citenamefont
  {Chakravarty}(2012)}]{chakraborty2012discrete}%
  \BibitemOpen
  \bibfield  {author} {\bibinfo {author} {\bibfnamefont {S.}~\bibnamefont
  {Chakraborty}}\ and\ \bibinfo {author} {\bibfnamefont {D.}~\bibnamefont
  {Chakravarty}},\ }\bibfield  {title} {\enquote {\bibinfo {title} {Discrete
  gamma distributions: Properties and parameter estimations},}\ }\href@noop {}
  {\bibfield  {journal} {\bibinfo  {journal} {Communications in
  Statistics-Theory and Methods}\ }\textbf {\bibinfo {volume} {41}},\ \bibinfo
  {pages} {3301--3324} (\bibinfo {year} {2012})}\BibitemShut {NoStop}%
\bibitem [{\citenamefont {Salgado-Garc\'{\i}a}\ and\ \citenamefont
  {Ugalde}(2016)}]{rsg2016symbolic}%
  \BibitemOpen
  \bibfield  {author} {\bibinfo {author} {\bibfnamefont {R.}~\bibnamefont
  {Salgado-Garc\'{\i}a}}\ and\ \bibinfo {author} {\bibfnamefont
  {E.}~\bibnamefont {Ugalde}},\ }\bibfield  {title} {\enquote {\bibinfo {title}
  {Symbolic complexity for nucleotide sequences: a sign of the genome
  structure},}\ }\href {http://stacks.iop.org/1751-8121/49/i=44/a=445601}
  {\bibfield  {journal} {\bibinfo  {journal} {J. of Phys. A: Math. and Theor.}\
  }\textbf {\bibinfo {volume} {49}},\ \bibinfo {pages} {445601} (\bibinfo
  {year} {2016})}\BibitemShut {NoStop}%
\bibitem [{\citenamefont {Gorman}\ and\ \citenamefont
  {Greene}(2008)}]{Gorman2008Visualizing}%
  \BibitemOpen
  \bibfield  {author} {\bibinfo {author} {\bibfnamefont {J.}~\bibnamefont
  {Gorman}}\ and\ \bibinfo {author} {\bibfnamefont {E.~C.}\ \bibnamefont
  {Greene}},\ }\bibfield  {title} {\enquote {\bibinfo {title} {Visualizing
  one-dimensional diffusion of proteins along dna},}\ }\href {\doibase
  10.1038/nsmb.1441} {\bibfield  {journal} {\bibinfo  {journal} {Nature
  Structural \& Molecular Biology}\ }\textbf {\bibinfo {volume} {15}},\
  \bibinfo {pages} {768--774} (\bibinfo {year} {2008})}\BibitemShut {NoStop}%
\bibitem [{\citenamefont {Mirny}\ \emph {et~al.}(2009)\citenamefont {Mirny},
  \citenamefont {Slutsky}, \citenamefont {Wunderlich}, \citenamefont {Tafvizi},
  \citenamefont {Leith},\ and\ \citenamefont {Kosmrlj}}]{Mirny2009How}%
  \BibitemOpen
  \bibfield  {author} {\bibinfo {author} {\bibfnamefont {L.}~\bibnamefont
  {Mirny}}, \bibinfo {author} {\bibfnamefont {M.}~\bibnamefont {Slutsky}},
  \bibinfo {author} {\bibfnamefont {Z.}~\bibnamefont {Wunderlich}}, \bibinfo
  {author} {\bibfnamefont {A.}~\bibnamefont {Tafvizi}}, \bibinfo {author}
  {\bibfnamefont {J.}~\bibnamefont {Leith}}, \ and\ \bibinfo {author}
  {\bibfnamefont {A.}~\bibnamefont {Kosmrlj}},\ }\bibfield  {title} {\enquote
  {\bibinfo {title} {How a protein searches for its site on {DNA}: the
  mechanism of facilitated diffusion},}\ }\href {\doibase
  10.1088/1751-8113/42/43/434013} {\bibfield  {journal} {\bibinfo  {journal}
  {Journal of Physics A: Mathematical and Theoretical}\ }\textbf {\bibinfo
  {volume} {42}},\ \bibinfo {pages} {434013} (\bibinfo {year}
  {2009})}\BibitemShut {NoStop}%
\bibitem [{\citenamefont {Salgado-Garc\'{\i}a}(2019)}]{rsg2019noise}%
  \BibitemOpen
  \bibfield  {author} {\bibinfo {author} {\bibfnamefont {R.}~\bibnamefont
  {Salgado-Garc\'{\i}a}},\ }\bibfield  {title} {\enquote {\bibinfo {title}
  {Noise-induced rectification in out-of-equilibrium structures},}\ }\href
  {\doibase 10.1103/PhysRevE.99.012128} {\bibfield  {journal} {\bibinfo
  {journal} {Phys. Rev. E}\ }\textbf {\bibinfo {volume} {99}},\ \bibinfo
  {pages} {012128} (\bibinfo {year} {2019})}\BibitemShut {NoStop}%
\bibitem [{\citenamefont {Cocho}\ \emph {et~al.}(2003)\citenamefont {Cocho},
  \citenamefont {Cruz}, \citenamefont {Mart\'{\i}nez-Mekler},\ and\
  \citenamefont {Salgado-Garc\'{\i}a}}]{cocho2003replication}%
  \BibitemOpen
  \bibfield  {author} {\bibinfo {author} {\bibfnamefont {G.}~\bibnamefont
  {Cocho}}, \bibinfo {author} {\bibfnamefont {A.}~\bibnamefont {Cruz}},
  \bibinfo {author} {\bibfnamefont {G.}~\bibnamefont {Mart\'{\i}nez-Mekler}}, \
  and\ \bibinfo {author} {\bibfnamefont {R.}~\bibnamefont
  {Salgado-Garc\'{\i}a}},\ }\bibfield  {title} {\enquote {\bibinfo {title}
  {Replication ratchets: polymer transport enhanced by complementarity},}\
  }\href@noop {} {\bibfield  {journal} {\bibinfo  {journal} {Physica A:
  Statistical Mechanics and its Applications}\ }\textbf {\bibinfo {volume}
  {327}},\ \bibinfo {pages} {151--156} (\bibinfo {year} {2003})}\BibitemShut
  {NoStop}%
\bibitem [{\citenamefont {Benson}\ \emph {et~al.}(1997)\citenamefont {Benson},
  \citenamefont {Boguski}, \citenamefont {Lipman},\ and\ \citenamefont
  {Ostell}}]{benson1997genbank}%
  \BibitemOpen
  \bibfield  {author} {\bibinfo {author} {\bibfnamefont {D.~A.}\ \bibnamefont
  {Benson}}, \bibinfo {author} {\bibfnamefont {M.~S.}\ \bibnamefont {Boguski}},
  \bibinfo {author} {\bibfnamefont {D.~J.}\ \bibnamefont {Lipman}}, \ and\
  \bibinfo {author} {\bibfnamefont {J.}~\bibnamefont {Ostell}},\ }\bibfield
  {title} {\enquote {\bibinfo {title} {Genbank},}\ }\href {\doibase
  10.1093/nar/25.1.1} {\bibfield  {journal} {\bibinfo  {journal} {Nucleic acids
  research}\ }\textbf {\bibinfo {volume} {25}},\ \bibinfo {pages} {1--6}
  (\bibinfo {year} {1997})}\BibitemShut {NoStop}%
\end{thebibliography}%

\end{document}